\documentclass[12pt]{iopart}

  \expandafter\let\csname equation*\endcsname\relax
  \expandafter\let\csname endequation*\endcsname\relax

\usepackage{graphicx,amssymb, amsmath}

\usepackage{isomath}

\usepackage{xr}

\let\vec\vectorsym
\let\tens\tensorsym

\newcommand{\uvec}[1]{{\vec{u}_{#1}}}
\newcommand{\intr}[1]{\bar{#1}}
\newcommand{\xma}[1]{{\tens{#1}_{\scriptscriptstyle\times}}}

\newcommand{\ff}{{\it ff}}

\newcommand{\diag}{{\rm diag}}
\newcommand{\com}{{\scriptscriptstyle{\mathbb{C}}}}

\newcommand{\eff}{{\it eff}}
\renewcommand\Re{\operatorname{Re}}
\renewcommand\Im{\operatorname{Im}}

\let\originalleft\left
\let\originalright\right
\renewcommand{\left}{\mathopen{}\mathclose\bgroup\originalleft}
\renewcommand{\right}{\aftergroup\egroup\originalright}

\newcommand{\JLTP}{{\it J. Low Temp. Phys. }}
\newcommand{\PC}{{\it Physica C }}
\newcommand{\IEEEas}{{\it IEEE Trans. Appl. Supercond. }}
\newcommand{\JSNM}{{\it J. Supercond. Nov. Magn. }}
\newcommand{\PRB}{{\it Phys. Rev. B }}
\newcommand{\APL}{{\it Appl. Phys. Lett. }}
\newcommand{\ZPB}{{\it Z. Phys. B }}
\newcommand{\NatMat}{{\it Nat. Mater. }}
\newcommand{\NatPhys}{{\it Nat. Phys. }}
\newcommand{\JETP}{{\it Soviet Phys.-JETP }}
\newcommand{\EPL}{{\it Europhys. Lett. }}
\newcommand{\IEEEmag}{{\it IEEE Trans. on Magnetics }}
\newcommand{\MPLB}{{\it Mod. Phys. Lett. B }}
\newcommand{\JS}{{\it J. Supercond. }}
\newcommand{\EPJB}{{\it Eur. Phys. J. B }}
\newcommand{\IJMPB}{{\it Int. J. Mod. Phys. B }}

\begin{document}

\title[d.c. and a.c. resistivity in anisotropic superconductors in tilted magnetic fields.]{Theory of measurements of d.c. and a.c. resistivity in anisotropic superconductors in tilted magnetic fields.}

\author{N. Pompeo}%
\address{Dipartimento Ingegneria, Universit\`a Roma Tre, Via della Vasca Navale 84, 00146 Roma, Italy}%
\ead{nicola.pompeo@uniroma3.it}

\begin{abstract}
The vortex dynamics of uniaxial anisotropic superconductors with arbitrary angles between the magnetic field, the applied current and the anisotropy axis is theoretically studied, by focusing on the models for electrical transport experiments in the linear regime. 
The vortex parameters, such as the viscous drag, the vortex mobility and the pinning constant (in the weak point pinning regime), together with the vortex motion resistivity, are derived in tensor form by considering the very different free flux flow and pinned Campbell regimes. The results are extended to high frequency regimes where additional effects like thermal depinning/creep take place.
The applicability to the various tensor quantities of the well-known scaling laws for the angular dependence on the field orientation is commented, illustrating when and with which  cautions the scaling approach can be used to discriminate between intrinsic and extrinsic effects.
It is shown that the experiments do not generally yield the intrinsic values of the vortex parameters and vortex resistivities. Explicit expressions relating measured and intrinsic quantities are given and their use exemplified in data analyses of angular measurements.
\end{abstract}


\maketitle

\section{Introduction}
Many superconductors of wide interest and recent discovery, such as iron-based superconductors  \cite{ironsup}, MgB$_2$ \cite{mgb2} and cuprate superconductors \cite{poole}, have in common an intrinsic material anisotropy, essentially uniaxial, arising from their crystal structure.
The material anisotropy has a profound impact, among the others, on the vortex dynamics and on the related pinning phenomena. Such properties have been much studied due to their importance both for unraveling the fundamental physics of the underlying superconductor and in view of technological applications \cite{anisThEx,coated1,coated2,JAP2009,APL2013p}. 
As an example, recently a great deal of effort have been devoted to the artificial tailoring of pinning on YBa$_2$Cu$_3$O$_{7-\delta}$ \cite{coated1,coated2,JAP2009}, through the introduction of defects of various geometries, which can introduce additional sources of anisotropy. 

I focus on the models for d.c. and a.c. electrical transport measurements in the linear regime in the mixed state, since this class of measurements is largely used in the study of vortex dynamics.
The interplay between the material anisotropy and the preferential direction introduced by the magnetic field $\vec{B}$ determines a non-straightforward relationship between the applied current density $\vec{J}$ and the corresponding electric field $\vec{E}$. Indeed, by applying with a arbitrary orientations $\vec{J}$ and $\vec{B}$, the vortices move under the effect of the Lorentz force $\vec{J}\times\uvec{B}\Phi_0$ and induce (Faraday's law) an electric field $\vec{E}$ which is in general not parallel to $\vec{J}$ even in isotropic superconductors. In anisotropic superconductors, additionally, in general vortices do not move parallel to the Lorentz force, further turning away $\vec{E}$ from the $\vec{J}$ direction.
As a consequence, the measured quantities, such as the flux flow, the Campbell and the high frequency resistivities (and their vortex counterparts, the vortex viscosity, the pinning constant and the complex viscosity), depend on the angles between $\vec{B}$, $\vec{J}$ and the anisotropy axis. Given the interest in studies with arbitrary orientations of both $\vec{B}$ and $\vec{J}$ \cite{scalingExp_2,caxisrho_lorentzXZ,aaxis}, and the versatility of the microwave techniques \cite{trunin} as tools for applying both in-plane and out-of-plane currents, a tensor representation and within it the accurate identification of the material intrinsic properties are needed.

Previous works addressed some aspects of the problem, such as: anisotropic flux flow in the pin-free d.c. regime \cite{HHTelectrodyn,HHTelectrodyn2,HHTtdgl,Ivlev1991,Genkin1989,haoclem}; pinning in non linear regimes in tilted fields, studied in the perspective of magnetization measurements \cite{brandttensor, klupsch}; two-dimensional anisotropic pinning with isotropic viscous drag and fixed magnetic field orientation \cite{Shklovskij}; coupling between anisotropic two-fluid currents and vortex motion, the latter described within an isotropic framework \cite{CCaniso,coffeyJLTP}.

In this work I propose a generalized treatment, centered on the force equation for the vortex motion, referring to uniaxial anisotropic superconductors in the mixed state.
Both the material anisotropy and pinning, the latter limited to weak random point pins only, are considered and studied in various regimes for arbitrary angles between $\vec{B}$, $\vec{J}$ and the anisotropy axis. I consider the purely dissipative free flux flow regime, the dissipationless pinned Campbell, and the high frequency regimes, where dissipation and pinning effects are comparable and additional phenomena like vortex creep become relevant \cite{BrandtModel,CCiso, creepCorbino, APL2007,JSNM2013a, SongPRB79}.
The goal is to provide the tensor representation of both the vortex parameters and the vortex motion resistivities and to relate the intrinsic quantities to the experimentally measured quantities, also by commenting on the applicability to the various tensors of the well-known scaling laws for the angular dependence on the field orientation \cite{klemm,scaling2,BGL,blatterone}. 

Moreover, in several examples the tensor expressions will be cast into expressions directly exploitable in the experiments and applied to analyze experimental data. 

This work is organized as follows: in section \ref{sec:electrodynamics_model}, the electrodynamics model is recalled; in section \ref{sec:rhoffi_scalinglaw} the intrinsic flux flow tensor expression and the angular scaling laws are briefly recalled; in section \ref{sec:fluxflowparameters} the vortex viscosity, vortex mobility and flux flow resistivity tensors are computed starting from the vortex force equation; in section \ref{sec:campbell} the treatment is extended to the a.c. Campbell regime, yielding the various pinning-related tensors; in section \ref{sec:highfrequency} high frequency regimes are studied, where both dissipation and pinning effects are taken into account, the latter including also thermal depinning/creep. Sections \ref{sec:exp_Campbell} and \ref{sec:rhovm_examples} are devoted to experimental aspects, providing examples of data analysis in the measurement of the pinning constant and of the high frequency resistivity, respectively. 

Throughout this manuscript, a vector is denoted as $\vec{A}=\uvec{A}A$, where $\uvec{A}$ and $A$ are its unit vector and modulus, respectively, while a tensor/matrix is denoted as $\tens{A}$. Moreover, ``$\diag(a_1, a_2,... a_n)$'' denotes a diagonal square matrix having diagonal scalar elements $a_1...a_n$.

\section{Vortex motion electrodynamics model}
\label{sec:electrodynamics_model}

Here I recall the model which will be used in this manuscript as a general framework to describe the superconductor electrodynamics response in the mixed state, and as such applied to the various regimes  studied. The starting point is the electrodynamics model proposed by Hao, Hu and Ting in their study of the d.c. flux flow resistivity in anisotropic superconductors \cite{HHTelectrodyn,HHTelectrodyn2}. The model holds in the linear regime in a homogeneous superconductor with an uniform magnetic field applied along a general direction, in the London limit. Vortices are assumed to be straight and rigid flux lines moving in a uniform current field density. This basic model does not take into account more complex phenomena such as helical instabilities of the vortex lines \cite{brandtone}, flux-line cutting effects \cite{gonzalez}, and breaking of vortex lines into pancakes in the extremely anisotropic, layered superconductors \cite{brandtone}.
Accordingly, I do not consider any electric field component $\parallel\vec{B}$ \cite{brandttensor}.

When vortices move with velocity $\vec{v}$, they induce an electric field $\vec{E}$ given by the Faraday's law \cite{EBv}: 
\begin{equation}
\label{eq:Ev}
\vec{E}=\vec{B}\times\vec{v}
\end{equation}
This vortex motion determines a transport current $\vec{J}_T$, related to $\vec{E}$ through the material conductivity tensor $\tens{\intr{\sigma}}$, so that, within the linear response theory \cite{HHTelectrodyn,HHTelectrodyn2}:
\begin{equation}
\label{eq:sigmaff_def}
  \vec{J}_T=\tens{\intr{\sigma}} \vec{E}
\end{equation}
The tensor $\tens{\intr{\rho}}=\left(\tens{\intr{\sigma}}\right)^{\!-1}$ is the \emph{material intrinsic} resistivity tensor:
\begin{equation}
\label{eq:rhoff_def}
  \vec{E}=\tens{\intr{\rho}}\vec{J}_T
\end{equation}
These tensors are a property of the superconductor material, they do not depend on the current intensity and orientation, and relate the current density $\vec{J}_T$ and the electric field $\vec{E}$ coupled by vortex motion. Their specific expressions depend on the considered vortex motion regime: explicit expressions in various regimes will be given in the following sections.

In experimental measurements, an externally imposed current $\vec{J}$ sets in motion the vortices  by exerting the Lorentz force (per unit length) $\vec{F}_L=\Phi_0 \vec{J}\times\uvec{B}$ on the individual vortices. Since $\vec{J}$ can be in principle arbitrarily oriented, it will be in general distinct from $\vec{J}_T$, which is instead constrained to specific orientations with respect to $\vec{B}$ (i.e. the condition $\tens{\intr{\rho}}\vec{J}_T\cdot\vec{B}=0$ holds).
The difference $\vec{J}_S=\vec{J}_T-\vec{J}$ is a supercurrent density $\vec{J}_S\parallel\vec{B}$ (i.e. $\vec{J}_S=J_S\uvec{B}$) \cite{HHTelectrodyn,HHTelectrodyn2} uncoupled with vortex motion. Since $\vec{J}\neq\vec{J}_T$, the \emph{experimentally measured} resistivity tensor $\tens{\rho}$, defined as:
\begin{equation}
\label{eq:rhoeff_def}
\vec{E}=\tens{\rho}\vec{J}
\end{equation}
is different from $\tens{\intr{\rho}}$. 
In particular, the widely used scalar resistivity $\rho^{(J)}$ measured along the direction of the applied current $\vec{J}$ is related to $\tens{\rho}$ as follows:
\begin{equation}
\label{eq:rhoJ}
\rho^{(J)}=\vec{J}\cdot\vec{E}/J^2=\uvec{J}\cdot\vec{E}/J=\left(\tens{\rho}\uvec{J}\right)\cdot\uvec{J}
\end{equation}
With a bit of algebra, the relation between the intrinsic resistivity tensor $\tens{\intr{\rho}}$  and the experimentally measured $\tens{\rho}$ can be worked out. For a diagonal $\tens{\intr{\rho}}$, one obtains:
\begin{equation}
\label{eq:rhopff}
\tens{\rho}=-\xma{B}\left(\frac{\left|\tens{\intr{\rho}}\right|\tens{\intr\rho}^{-1}}{\left(\tens{\intr{\rho}}\uvec{B}\right)\cdot\uvec{B}}\right)\xma{B}
\end{equation}
where $\xma{B}=\tens{\epsilon}_{ijk}\uvec{B}$, being $\tens{\epsilon}_{ijk}$ the permutation (Levi-Civita) tensor \cite{Itskov}.
It is worth stressing that while the elements of the tensor $\tens{\rho}$ can be (by definition) directly measured by properly choosing the direction of the applied $\vec{J}$ and of the measured $\vec{E}$ component, the same does not hold in general for the elements of the intrinsic $\tens{\intr{\rho}}$ tensor which, on the other hand, is the physical quantity of interest. 
Therefore it becomes necessary to determine $\tens{\intr{\rho}}$ from measurements of $\tens{\rho}$ by paying attention to the removal of the additional ``spurious'' contributions present in the latter, as it will be illustrated in the remaining part of this paper and in particular in the sections \ref{sec:exp_Campbell} and \ref{sec:rhovm_examples}.

\section{Flux flow regime}

\subsection{Intrinsic flux flow resistivity tensor and angular scaling laws}
\label{sec:rhoffi_scalinglaw}

\begin{figure}[ht]
\centerline{\includegraphics[width=4cm]{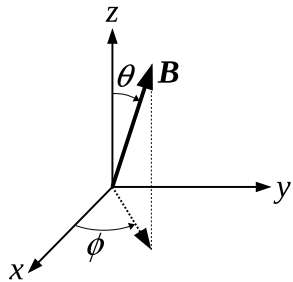}}
  \caption{ Principal frame of reference. The magnetic induction field $\vec{B}$ is also depicted, applied along a general direction at the polar $\theta$ and azimuthal $\phi$ angles.}
\label{fig:ref}
\end{figure}

The explicit expression for the intrinsic tensor $\tens{\intr{\rho}}_\ff$ in the flux flow regime has been previously derived \cite{HHTtdgl,Ivlev1991,Genkin1989,haoclem}. I follow reference \cite{HHTtdgl} which, working within the Time-Dependent Ginzburg-Landau (TDGL) theory (${B}\lesssim {B}_{c2}$), accounts for both ohmic losses and order parameter relaxation.
I first specify the frame of reference. The crystallographic axes are taken as the coordinate axes of a Cartesian frame of reference, having coordinate unit vectors $\uvec{x}$, $\uvec{y}$ and $\uvec{z}$, so that $x\equiv a$, $y\equiv b$ and $z\equiv c$, being the latter the axis of the uniaxial anisotropy. The chosen frame of reference, together with a magnetic induction field vector with general orientation $\vec{B}=B\uvec{B}$, is  depicted in figure \ref{fig:ref}.
In this frame of reference the phenomenological electronic mass tensor \cite{klemm,kogan}, which can be used to describe in the London limit the material anisotropy, is diagonal:
\begin{equation}
\label{eq:masstensor}
\diag(m_{ab},m_{ab},m_{c})=m\tens{M}
\end{equation}
having defined the in-plane mass $m_{ab}=m$, the out-of-plane mass $m_c$, the anisotropy factor $\gamma^2=m_c/m_{ab}$ and $\tens{M}=\diag(1,1,\gamma^2)$.
The mass tensor $\tens{M}$ contains all the information concerning the material anisotropy, since the only source of anisotropy that will be considered is the effective mass of the charge carriers (this implies, for example, that the possible anisotropy of the scattering time of the normal carriers is neglected).
By neglecting the Hall contribution and assuming the same anisotropy axes for the normal resistivity and mass tensors $\tens{\rho}_n=\rho_{n,11}\tens{M}$ (i.e. the mass tensor is the only source of anisotropy also for the normal state), the tensors $\tens{\intr{\rho}}_\ff$ and $\tens{\intr{\sigma}}_\ff$ are \emph{diagonal} in the stated frame of reference:
\begin{equation}
\label{eq:rhoff_tensor1}
\tens{\intr{\rho}}_\ff(B, \theta)=\diag(\intr\rho_{\ff,11}, \intr\rho_{\ff,11}, \intr\rho_{\ff,33})
\end{equation}
where $\intr\rho_{\ff,ii}=\intr\rho_{\ff,ii}(B, \theta)$. Neglecting the weak field dependence of $\tens{\rho}_n$ one obtains \cite{HHTelectrodyn,HHTtdgl}: 
\begin{equation}
\label{eq:rhoff_element}
\intr\rho_{\ff,ii}(B,\theta)/\rho_{n,ii}=\mathcal{F}(B/B_{c2}(\theta))
\end{equation}
which shows that all the three tensor elements of $\tens{\intr{\rho}}_\ff$ share the same field dependence through a common function $\mathcal{F}(B/B_{c2})$ and that the common field dependence is consistent with the scaling laws \cite{klemm,scaling2,BGL,blatterone}, widely observed and experimentally studied \cite{scalingExp,scalingExp_2}. The Blatter-Geshkenbein-Larkin (BGL) formalization of the scaling law \cite{BGL,blatterone} states that in the London approximation, a thermodynamic or intrinsic transport property $q$ of a uniaxially anisotropic superconductor depends on the applied field intensity $B$ and angle $\theta$ through the scaled field $B\epsilon(\theta)$ as $q(B,\theta)=s_q q^{iso}(B\epsilon(\theta))$, where $\epsilon(\theta)=(\cos^2\theta+\gamma^{-2}\sin^2\theta)^{1/2}$ is the angular-dependent anisotropy parameter, $q^{iso}(B)$ is the field-dependent quantity in the equivalent isotropic superconductor, and the scaling factor is typically $s_q=\gamma^\alpha \epsilon^\beta(\theta)$.
From this rule one obtains readily $q(B,\theta)=\epsilon^\beta(\theta) q(B\epsilon(\theta),0)$. 
For $\beta=0$, as it happens for $\intr\rho_\ff$, one has $q(B,\theta)=q(B\epsilon(\theta),0)$, 
i.e. the anisotropic physical quantity in the mixed state depends on the $B/B_{c2}(\theta)$ ratio only ($B_{c2}(\theta)=B_{c2}(0)/\epsilon(\theta)$).
The scaling approach holds also in presence of point pins, but not with other pin geometries like extended defects. In this work the eventual adherence to the scaling law of the various anisotropic quantities will be commented, given its usefulness as a tool to identify the intrinsic (charge carrier mass) material anisotropy, including also the effect of point pins, as opposed to other eventual extrinsic sources of anisotropy, such as extended defects.

Equation \eref{eq:rhoff_tensor1} can be rewritten as:
\begin{equation}
\label{eq:rhoff_tensor2}
\tens{\intr\rho}_\ff(B,\theta)=\intr\rho_{\ff,11}(B,\theta)\tens{M}={\intr\rho}_{\ff,11}(B\epsilon(\theta),0)\tens{M}
\end{equation}
The above equation highlights the important property that the field magnitude and angular dependence of the whole intrinsic flux flow resistivity tensor can be represented by a single scalar function, namely the element $\intr\rho_{\ff,11}(B,\theta)$, which adheres to the angular scaling law.
Similarly, the conductivity tensor is $\tens{\intr\sigma}_\ff(B,\theta)=\intr\sigma_{\ff,11}(B,\theta)\tens{M}^{-1}$.

In the following, for ease of notation the field magnitude and angular dependence will be explicitly written only in the equations reporting the main results.

\subsection{Vortex parameters in the flux flow regime}
\label{sec:fluxflowparameters}
In this section the well-known vortex force equation \cite{GR,Golo}, involving the balance of forces (per unit length) acting on an individual vortex, is studied in the regime of pure flux flow:
\begin{equation}
\label{eq:force1}
\tens{\eta}\vec{v}=\Phi_0 \vec{J}\times\uvec{B}
\end{equation}
where $-\tens{\eta}\vec{v}$ is the viscous drag force and $\tens{\eta}$ is the viscous drag, also known as viscosity, tensor.
Experimentally, this regime can be obtained when pinning is very low \cite{kunchurPRB82}, with d.c. currents sufficiently large to overcome the pinning forces \cite{highCurrentFF}, or with a.c. currents at high enough frequencies \cite{GR}. The issue of the true d.c. current path is of paramount importance by itself, and it has to be solved separately \cite{esposito}. 
In any case, it has to be kept in mind that the actual realization of the flux flow regime requires specific conditions and care \cite{GR,kunchurPRB82,highCurrentFF,unstableFF}.

The following computations yield a generalized formulation which provides (i) the vortex viscosity  $\tens{\eta}$ and the related vortex mobility $\tens{\mu}_v$ tensors; (ii) the relationship between $\tens{\eta}$ and the flux flow resistivity tensors $\tens{\intr\rho}_\ff$ and $\tens{\rho}_\ff$; (iii) the basis for the extension of the tensor-based treatment to a.c. regimes with pinning (sections \ref{sec:campbell} and \ref{sec:highfrequency}).

\subsubsection{Vortex viscosity tensor.}
\label{sec:viscosity}

Starting from the force equation \eref{eq:force1} and working within the above recalled electrodynamics model, with straightforward calculations one obtains:
\begin{equation}
\label{eq:eta}
\tens{\eta}(B, \theta,\phi)=-\Phi_0 B\xma{B}\tens{\intr\sigma_\ff}\xma{B}=\Phi_0 B\intr\sigma_{\ff,11}(B,\theta)\tens{M}_B(\theta,\phi)
\end{equation}
where $\tens{M}_B(\theta,\phi)=-\xma{B}\tens{M}^{-1}\xma{B}$ has been introduced for the sake of compactness.
It can be noted that the viscosity tensor $\tens{\eta}$ does not obey the angular scaling law and that it depends also on $\phi$, even if the superconductor is uniaxially anisotropic, because of the Faraday-Lorentz contribution included in $\tens{M}_B(\theta,\phi)$.
For symmetry of notation, an ``intrinsic'' viscosity tensor can be introduced as follows:
\begin{equation}
\label{eq:eta0i}
\tens{\intr\eta}(B,\theta)=\Phi_0 B \tens{\intr{\sigma}}_\ff(B,\theta)=\frac{\intr\eta_{11}(B\epsilon(\theta),0)}{\epsilon(\theta)}\tens{M}^{-1}\\
\end{equation}
In the last member, it is shown that $\tens{\intr\eta}$ satisfies the angular scaling law with $s_\eta\propto \epsilon^{-1}(\theta)$.
Equation \eref{eq:eta} can be thus rewritten as:
\begin{equation}
\label{eq:eta0}
\tens{\eta}(B,\theta,\phi)=-\xma{B}(\theta,\phi)\tens{\intr\eta}(B,\theta)\xma{B}(\theta,\phi)
\end{equation}
Contrary to $\tens{\intr\eta}$ and $\tens{\intr{\sigma}}_\ff$, the viscosity tensor $\tens{\eta}$ is not diagonal. 
Nevertheless it can be diagonalized by finding its eigenvectors $\vec{v}_e$ $(\uvec{v}_{e1}={(\uvec{B}\times\uvec{z})}/{|\uvec{B}\times\uvec{z}|}$, $\uvec{v}_{e2}=\uvec{B}\!\times\!\uvec{v}_{e1}$, $\uvec{v}_{e3}=\uvec{B})$ and corresponding eigenvalues $(\eta_{e,i}=\intr\eta_{11}\epsilon^2, \intr\eta_{11}, 0)$ so that in the new frame of reference $\tens{T}=[\uvec{v}_{e1}\;\uvec{v}_{e2}\;\uvec{v}_{e3}]$ the viscosity tensor $\tens{\eta'}$ is \cite{Itskov}:
\begin{equation}
\label{eq:roteta}
\tens{\eta'}=\tens{T}^{-1}\tens{\eta}\tens{T}=\intr\eta_{11}\diag(\epsilon^2,1,0)
\end{equation}
As a particular case, it can be checked that the present expression of the tensor $\tens{\eta}$ yields the same (eigen)values of the viscosity along the principal axes as those reported in \cite{haoclem}, computed within the Bardeen\textendash{}Stephen \cite{BS} model.

\begin{figure}[ht]
\centerline{\includegraphics[width=8.5cm]{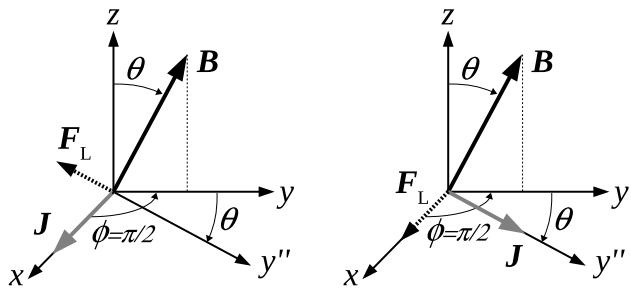}}
  \caption{ Field-current configurations related to the viscosity eigenvectors (see text).}
\label{fig:fieldcurconf}
\end{figure}

Since $\uvec{v}_{e1}$ and $\uvec{v}_{e2}$ are  $\perp\uvec{B}$, the force equation \eref{eq:force1} implies that for both of them there exists a current direction $\uvec{J}$ yielding a right-handed orthogonal basis $[\uvec{J},\uvec{B},\uvec{v}_{e}]$ in which $\vec{E}\parallel\vec{J}$, so that the resistivity tensor is effectively represented by a single scalar value. 
Having taken $\vec{B}$ in the $y$-$z$ plane for ease of notation and without loss of generality, these current directions are $\uvec{J1}=\uvec{y''}=\uvec{B}\times\uvec{x}$ and $\uvec{J2}=\uvec{x}$, and are drawn in figure \ref{fig:fieldcurconf}, in the right and left panel respectively.
The configuration on the left panel of figure \ref{fig:fieldcurconf} is the well-known and widely used ``maximum Lorentz force'' setup. The other one is, among other possible geometries \cite{scalingExp_2,caxisrho_lorentzXZ}, seldom used in the experiments, because it requires samples grown with the c-axis parallel to the sample surface \cite{aaxis}. Nevertheless, its usefulness will become apparent through its use in the study of the Campbell pinning regime proposed in section \ref{sec:campbell}.

\subsubsection{Vortex mobility tensor.}
\label{sec:mobility}
The force equation \eref{eq:force1} can be written in terms of the vortex mobility tensor $\tens{\mu}_v$ as:
\begin{equation}
\label{eq:force2}
\vec{v}=\tens{\mu}_v\Phi_0 \vec{J}\times\uvec{B}
\end{equation}
Since the vortex viscosity tensor has rank=2 (equation \eref{eq:roteta}), $\tens{\mu}_v$ cannot straightforwardly be derived as the inverse of $\tens{\eta}$, as the relation between the corresponding scalar quantities would suggest.
Nevertheless, in the plane $\perp B$, $\tens{\mu}_v$ is the inverse bijection of $\tens{\eta}$ whereas the vortex mobility for the (physically not existing) Lorentz force component $\parallel\uvec{B}$ can be safely set $=0$.
Hence:
\begin{equation}
\label{eq:mu_expr}
\tens{\mu'}_{v}=\left(\intr\eta_{11}\right)^{\!-1}\diag(\epsilon^{-2},1,0)
,\tens{\mu}_{v}=\tens{T}\tens{\mu'}_{v}\tens{T}^{-1}
\end{equation}

\subsubsection{Relation between vortex viscosity and measured flux flow resistivity tensors}
\label{sec:rhoff}
The flux flow resistivity tensor $\tens{\rho}_\ff$ can be now expressed in terms of the vortex mobility $\tens{\mu}_{v}$. Starting from equation \eref{eq:force2} one obtains, after a little algebra:
\begin{equation}
 \label{eq:rhopff2}
\tens{\rho}_\ff=-\xma{B}\tens{\mu}_v\Phi_0B\xma{B}
\end{equation}
It is of paramount importance that equation \eref{eq:rhopff2} represents the \emph{measured}, \emph{apparent}, flux flow resistivity tensor, \emph{different} from the material intrinsic flux flow resistivity tensor. The relation between the material $\tens{\intr{\rho}}_\ff$, the measured $\tens{\rho}_\ff$ and the vortex viscosity tensor $\tens\eta$ can be made explicit resorting to equation \eref{eq:mu_expr} and considering that $\tens{T}^{-1}\xma{B}\tens{T}=\xma{z}$:
\begin{equation}
\label{eq:rhopff2_1}
\tens{\rho}_\ff=\frac{\Phi_0B\tens{\eta}}{\left[\intr\eta_{11}\epsilon\right]^2}
\end{equation}
Using equations \eref{eq:eta}\textendash{}\eref{eq:eta0} the role of $\tens{\intr\rho}_\ff$ emerges:
\begin{equation}
\label{eq:rhoff_uni}
\tens{\rho}_\ff(B,\theta,\phi)=\intr\rho_{\ff,11}(B,\theta)\frac{\tens{M}_B(\theta,\phi)}{\epsilon^2(\theta)}
\end{equation}
It can be checked that the above equation is equivalent to equation \eref{eq:rhopff}, written with $\tens{\rho}=\tens{\rho}_\ff$. 
It can be noted that $\tens{\rho}_\ff$, like the viscosity tensor $\tens{\eta}$ of equation \eref{eq:eta} and contrary to $\tens{\intr\rho}_\ff$, does not obey the angular scaling law. 

The above expressions relating the measured resistivity tensor to the vortex parameters are an important result of this paper since they enable the computations performed in the following sections, devoted to the study of a.c. regimes. 

\section{Campbell regime}
\label{sec:campbell}

A totally different limit with respect to flux flow is given by the pinning regime. There, the d.c. response is zero. However, the a.c. resistivity is well measurable (\cite{Golo} and references therein). To my knowledge, no complete treatment of the anisotropic resistivity in the pinning regime exists. In this section I consider the vortex motion under a.c. currents flowing in anisotropic superconductors with random point pinning centers, described in the weak collective pinning regime \cite{blatterone}. In particular, I focus on the well-known Campbell regime, in which the pinning action on vortices dominates over the dissipative viscous drag and the pinning force is proportional to the small displacement $\Delta\vec{r}$ of the vortices from the pinning centers. The above two conditions are achieved respectively for frequencies smaller than the so-called pinning frequency (of the order of a few MHz in conventional superconductors \cite{GR} and of several GHz in high-$T_c$ superconductors \cite{APL2007, Golo, Wu1995, 
omegap_HTS,JSNM2006}, MgB$_2$ \cite{mgb2_omegap},  low-$T_c$ thin films \cite{SongPRB79,lowTc_omegap} and more complex superconducting heterostructures \cite{creepCorbino,JSNM2013a,trilayers}) and for currents small enough to ensure the validity of the linear regime. Considering only point pins (i.e. zero-dimensional pinning centers) ensures that no further preferential directions are introduced in the superconducting system, in contrast with what happens when extended (e.g. linear or planar) defects are present.

The starting point is the force equation written in the sinusoidal regime $e^{\rmi\omega t}$, including the pinning force and neglecting the viscous drag (hence neglecting losses):
\begin{equation}
\label{eq:force3}
\frac{1}{\rmi\omega}\tens{k}_p\vec{v}=\Phi_0 \vec{J}\times\uvec{B}
\end{equation}
where $\tens{k}_p$ is the pinning constant (also called Labusch parameter) tensor, $\vec{v}=\rmi\omega\Delta\vec{r}$ and $\vec{J}$ is a (low-frequency and small intensity) a.c. current. No creep phenomena are considered.

The corresponding so-called Campbell resistivity is purely imaginary and written as, in isotropic superconductors:
\begin{equation}
\label{eq:rhoC}
  \rho_C=\omega\frac{\Phi_0 B}{k_p}=\omega\mu_0\lambda_C^2
\end{equation}
where $\lambda_C$ is the Campbell penetration depth \cite{Campbellpenetration}. 

Going back to anisotropic superconductors, it is evident that the force equation \eref{eq:force3} in the Campbell regime is formally equivalent to the force equation \eref{eq:force1} written for the pure flux flow regime.
Hence, by applying the general electrodynamics model recalled in section \ref{sec:electrodynamics_model} and exploiting the subsequent computations of section \ref{sec:fluxflowparameters}, it is straightforward to introduce the series of tensors $[\tens{k}_p/(\rmi\omega), \tens{\intr k}_p/(\rmi\omega), -\rmi\tens{\intr\sigma}_C, \rmi\tens{\intr\rho}_C, \rmi\tens{\rho}_C]$ which are analogous to the already studied  $[\tens{\eta}, \tens{\intr\eta}, \tens{\intr{\sigma}}_\ff, \tens{\intr{\rho}}_\ff, \tens{\rho}_\ff]$. Therefore the following expressions can be obtained:
\begin{subequations}
\label{eq:pinningtensors}
\begin{align}
\label{eq:kptensor}
\tens{k}_p&=-\xma{B}\tens{\intr k}_p\xma{B} \\
\label{eq:kptensori}
\tens{\intr k}_p&=\omega\Phi_0B\tens{\intr\sigma}_{C}\\
\label{eq:rho_Ci}
\tens{\intr\rho}_C&=\left(\tens{\intr\sigma}_{C}\right)^{-1} \\
\label{eq:rho_C}
\tens{\rho}_{C}&=-\xma{B}\left(\frac{\left|\tens{\intr\rho}_C\right|\tens{\intr\rho}_C^{-1}}{\left(\tens{\intr\rho}_C\uvec{B}\right)\cdot\uvec{B}}\right)\xma{B}
\end{align}
\end{subequations}
where the latter expression holds for a diagonal $\tens{\intr\rho}_C$, as it will be taken in the following.
The above equations do not give a complete model in the Campbell regime, since an explicit expression for the Campbell tensor $\tens{\intr\rho}_C$ of the material is still needed. 
Whereas in the flux flow regime the computation of $\tens{\intr{\rho}}_\ff$ was addressed by the TDGL treatment \cite{HHTtdgl}, the determination of the tensor $\tens{\intr\rho}_C$ will be addressed in the following section.

\subsection{The Campbell resistivity tensor elements}
\label{sec:campbellelements}

I assume that $\tens{\intr\rho}_C$ is diagonal like $\tens{\intr{\rho}}_\ff$: since it can be easily checked that, when $\uvec{B}$ is parallel to a coordinate axis, the pinning tensor $\tens{k}_p$ is diagonal if and only if $\tens{\intr{\rho}}_C$ is diagonal, this ensures that choosing $\vec{B}$ and $\vec{J}$ along two coordinate axes, the corresponding pinning force $\vec{F}_p$ (and vortex velocity) will be parallel to the third coordinate axis, as it is reasonable to expect. 

This assumption leaves the diagonal elements, $\intr\rho_{C,11}(\theta)$ and $\intr\rho_{C,33}(\theta)$ within the uniaxial anisotropy, to be determined.
It can be easily shown that $\intr\rho_{C,11}(\theta)=\rho_{C}^{(x)}(\theta)$, which is the resistivity that can be experimentally measured with $\vec{J}\parallel\uvec{x}$ (equation \eref{eq:rhoJ}) and $\vec{B}(\theta,\phi=\pi/2)\in y$-$z$ plane (i.e. the well-known ``maximum Lorentz force'' configuration). On the other hand, $\intr\rho_{C,33}(\theta)$ is not directly accessible through real experimental current-field configurations (as it can be shown with a little algebra exploiting the expressions which relate $\tens{\rho}$ to $\tens{\intr\rho}$). Therefore it must be {\em indirectly} determined through coupled, complementary measurements. 
I choose $\rho_{C}^{(x)}(\theta)$ (already used for the above determination of $\intr\rho_{C,11}(\theta)$) and $\rho_{C}^{(y'')}(\theta)$, related to the two field-current configurations commented in section \ref{sec:viscosity} and depicted in figure \ref{fig:fieldcurconf}. 
Hence one obtains:
\begin{subequations}
\label{eq:rhoC_i}
\begin{align}
\label{eq:rhoC_i1}
  \intr\rho_{C,11}(B,\theta)&=\rho_{C}^{(x)}(B,\theta) \\
  \intr\rho_{C,33}(B,\theta)&=\frac{\rho_{C}^{(x)}(B,\theta)\sin^2\theta}{\frac{\rho_{C}^{(x)}(B,\theta)}{\rho_{C}^{(y'')}(B,\theta)}-\cos^2\theta}
\end{align}
\end{subequations}
According to section \ref{sec:viscosity}, both the field-current configurations ``(x)'' and ``($y''$)'' correspond to eigenvectors of $\vec{v}$, and hence of $\Delta\vec{r}\parallel\vec{v}$. The pinning force $-\tens{k}_p\Delta\vec{r}=-k_{p,e}\Delta\vec{r}$ is then completely described by a single scalar value, the eigenvalue of the pinning constant $k_{p,e}$ (denoted $k_p^{(x)}$ and $k_p^{(y'')}$ for the two configurations, respectively). 
The definition \eref{eq:rhoC} allows to write down the following:
\begin{equation}
\label{eq:rhoC_xy}
  \rho_{C}^{(x|y'')}(B,\theta)=\omega\frac{\Phi_0 B}{k_p^{(x|y'')}(B,\theta)}
\end{equation}
The explicit expressions of the pinning constants are determined as follows. 

\subsection{The pinning constant tensor elements}

By simple physical arguments \cite{koshelev}, for an isotropic superconductor one can evaluate the pinning constant $k_p$ by equating the maximum pinning force $k_p r_{pin}$ acting on a vortex, where $r_{pin}$ denotes the action range of the pinning centers, with the maximum Lorentz force $J_c\Phi_0$ exerted when the current equals the critical current density. Since for core-pinning $r_{pin}\sim\xi$, i.e. the coherence length defining the radius of the vortex core, one can write:
\begin{equation}
\label{eq:kp_Jc}
  k_p=c\frac{\Phi_0 J_c}{\xi}
\end{equation}
where $c\sim 1$. 

For anisotropic superconductors, equation \eref{eq:kp_Jc} must be specialized for the two current configurations ``(x)'' and ``($y''$)'' used in equation \eref{eq:rhoC_xy}. 
The pinning constant $k_p^{(x)}$ [$k_p^{(y'')}$] describes the pinning action on vortices moving along the $y''$ [$x$] direction under the action of a current $\parallel x$ [ $\parallel y''$]: hence the critical current density $J_c^{(x)}\parallel x$ [$J_c^{(y'')}\parallel y''$], and $r_{pin}\sim\xi^{(y'')}$ [$r_{pin}\sim\xi^{(x)}$], i.e. the coherence length along the direction $y''$ [$x$] of the vortex movement. Therefore:
\begin{equation}
\label{eq:kp_Jc_xy}
  k_p^{(x|y'')}(B,\theta)=c\frac{\Phi_0 J_c^{(x|y'')}(B,\theta)}{\xi^{(y''|x)}(\theta)}\\
\end{equation}
One should note that $k_p^{(y'')}$, despite being referred to a current-field configuration of difficult realization in the experiments, is related to quantities ($J_c^{(y'')}$ and $\xi^{(x)}$) which can be expressed within available theories. In this sense, equation \eref{eq:kp_Jc_xy} gives a tool for subsequent elaborations. Indeed, the above equations can be further developed using the BGL scaling law results \cite{blatterone}. The current densities $J_c^{(x)}$ and $J_c^{(y'')}$ for point-pinning in the single vortex and small-bundle pinning regime (the scaling theory does not describe the large bundle pinning regime, appearing in the highest field and temperature regions) are given by the following scaling expressions \cite{blatterone}:
\begin{subequations}
\label{eq:scalingJc}
\begin{align}
J_c^{(x)}(B,\theta)&=\gamma^{\frac{2}{3}}J_c^{iso}(B\epsilon(\theta))\\
  J_c^{(y'')}(B,\theta)&=\gamma^{\frac{2}{3}}\epsilon(\theta)J_c^{iso}(B\epsilon(\theta))
\end{align}
\end{subequations}
where the full expression of $J_c^{iso}(B)$ is reported in reference \cite{blatterone} for the different pinning regimes.
It is worth stressing that the critical current $J_c^{iso}(B)$ (and the corresponding $k_p^{iso}(B)=c{\Phi_0 J_c^{iso}(B)}/{\xi^{iso}}$) for the equivalent isotropic superconductor is not angle-dependent because point pins only are considered, whereas extended pinning centers would introduce preferential directions and therefore angle-dependent quantities even in an isotropic superconductor.

Geometrically, the quantities $\xi^{(y'')}(\theta)$ and $\xi^{(x)}(\theta)$ represent the maximum distances between the displaced vortex and the point pin \cite{blatterone}: 
\begin{subequations}
\label{eq:scalingxi}
\begin{align}
  \xi^{(y'')}(\theta)&=\epsilon(\theta)\xi\\
  \xi^{(x)}(\theta)&=\xi 
\end{align}
\end{subequations}
where $\xi=\xi^{iso}$ denotes the in-plane coherence length. 

By using equations \eref{eq:scalingJc} and \eref{eq:scalingxi} in equation \eref{eq:kp_Jc_xy}, one can write:
\begin{subequations}
\label{eq:kp_Jc_xy2}
\begin{align}
\label{eq:kp_Jc_xy2a}
  k_p^{(x)}(B,\theta)&=s_{k_p^{(x)}} k_p^{iso}(B\epsilon(\theta)) \text{, } s_{k_p^{(x)}}=\gamma^{\frac{2}{3}}\epsilon^{-1}(\theta)\\
  k_p^{(y'')}(B,\theta)&=s_{k_p^{(y'')}} k_p^{iso}(B\epsilon(\theta)) \text{, } s_{k_p^{(y'')}}=\gamma^{\frac{2}{3}}\epsilon(\theta)
\end{align}
\end{subequations}
It is worth stressing that, thanks to equation \eref{eq:rhoC_i1}:
\begin{equation}
\label{eq:kp11}
  \intr k_{p,11}(B, \theta)=k_p^{(x)}(B, \theta)
\end{equation}
hence the in-plane pinning constant $\intr k_{p,11}$ obeys the scaling law given by equation \eref{eq:kp_Jc_xy2a}.

\subsection{The Campbell regime tensors}

Using equations \eref{eq:rhoC_i}, \eref{eq:rhoC_xy} and \eref{eq:kp_Jc_xy2}, one obtains:
\begin{subequations}
\label{eq:rhoCscaling}
\begin{align}
  \intr\rho_{C,11}(B,\theta)&=s_{\rho_{C,11}} \rho_{C}^{iso}(B\epsilon(\theta)) \text{, } s_{\rho_{C,11}}=\gamma^{-\frac{2}{3}}\\
\intr\rho_{C,33}(B,\theta)&=\intr\rho_{C,11}(B,\theta)\gamma^2
\end{align}
\end{subequations}
Therefore, similarly to $\tens{\intr{\rho}}_\ff$ of equation \eref{eq:rhoff_tensor2}, one can write: 
\begin{equation}
\label{eq:rhoC_tensor}
  \tens{\intr\rho}_{C}(B,\theta)=\intr\rho_{C,11}(B,\theta)\tens{M}={\intr\rho}_{C,11}(B\epsilon(\theta),0)\tens{M}
\end{equation}
This is an important result of this paper: the explicit expression of $\tens{\intr\rho}_{C}$ (and the related $\tens{\intr k}_p$ given below) for point pinning has been obtained, showing that the anisotropy of the pinning tensors and the flux flow tensors is the same and is completely described by the mass anisotropy tensor. Moreover, both $\tens{\intr\rho}_\ff$ and $\tens{\intr\rho}_C$ satisfy the same angular scaling law $\tens{\intr\rho}_{\ff|C}(B,\theta)={\intr\rho}_{\ff|C,11}(B\epsilon(\theta),0)\tens{M}$. This result is very important also in the study of high frequency regimes (sections \ref{sec:highfrequency} and \ref{sec:rhovm_examples}) where both losses and pinning phenomena are equally relevant.

Lastly, recalling equation \eref{eq:kptensori}, the ``intrinsic'' pinning constant tensor can be written down:
\begin{equation}
\label{eq:kptensor_Bdep}
\tens{\intr k}_p(B,\theta)=\intr k_{p,11}(B,\theta)\tens{M}^{-1}=\frac{\intr k_{p,11}(B\epsilon(\theta),0)}{\epsilon(\theta)}\tens{M}^{-1}\\
\end{equation}
where $\intr k_{p,11}=\omega\Phi_0 B/\rho_{C,11}$
and where the last member highlights the scaling law satisfied by $\tens{\intr k}_p$.
Finally, the experimentally measured pinning constant tensor is: 
\begin{equation}
\tens{k}_p(B,\theta)=\intr k_{p,11}(B,\theta)\tens{M}_B(\theta,\phi)
\end{equation}
and it does not scale because of the additional contribution given by $\tens{M}_B(\theta,\phi)$.

\section{Application to experiment: a.c. linear susceptibility measurements analysis}
\label{sec:exp_Campbell}

In this section I provide an example showing the additional information which can be gained using the results of the present work by performing an analysis of experimental data.
Various techniques can be used to explore the Campbell regime, including inductance measurements \cite{inductive}, a.c. linear susceptibility measurements \cite{suscept} and the vibrating reed technique \cite{vibratingreed}. The pinning constant can be also determined through microwave measurements \cite{APL2007,JSNM2006,more_kp}, which requires more extended models dealt with in section \ref{sec:highfrequency}.
In the following, I focus on a.c. linear susceptibility measurements on YBa$_2$Cu$_3$O$_{7-\delta}$ samples in the mixed state, performed by varying both the direction and the intensity of the applied d.c. magnetic field. The source of the experimental data is reference \cite{Pasquini}, where full details about the experiment can be found. The measured squared real part of the penetration depth $\lambda^2_R=\lambda^2_L+\lambda_C^2$ ($\lambda_L$ is the London penetration depth) is related to the squared Campbell penetration depth $\lambda_C^2$ which, given the geometry of the experiment, is $\lambda_{C,11}\propto k_{p,11}^{-1}$.
The examined sample are two twinned YBa$_2$Cu$_3$O$_{7-\delta}$ single crystals, one irradiated at 30$^\circ$ with respect the $c$-axis in order to create columnar defects capable of reinforcing the pinning properties and one, virgin, used as reference.
As a consequence, four sources of pinning are expected: point pins and three types of correlated pinning centers, namely the twins along the $c$-axis direction, the $a$-$b$ planes and the columnar defects at 30$^\circ$ (in the irradiated sample only). It was concluded that pinning was stronger with the field aligned with columnar defects and, from a qualitative analysis, that an enhancement of pinning existed even far from the track direction. I show in the following that such findings can be put on solid quantitative grounds by exploiting the present model. In particular, I will use the model developed in the previous section to remove the anisotropic response due to the material anisotropy and to point pins, which can give rise to a significant angle-dependent contribution possibly obscuring the angle-dependent contributions arising from correlated defects.

Here I consider the data taken from figure 5 of reference \cite{Pasquini} for the irradiated sample only, where $\lambda^2_R(B)$ measured at various angles $\theta$ at T=$90.5\;$K is reported. In order to extract $k_{p,11}(B,\theta)$, I extrapolate $\lambda^2_R(B\rightarrow 0)$, take it as an evaluation of $\lambda^2_L$ and, by neglecting pair-breaking effects, estimate $\lambda_C^2(B)=\lambda^2_R(B)-\lambda^2_R(B\rightarrow 0)$. Then the pinning constant $k_{p,11}\propto B/\lambda^2_C(B)$ is obtained, using equation \eref{eq:rhoC} with $B=\mu_0H$. By dividing by $k_{p,11}^{max}$ (i.e. the maximum value attained by $k_{p,11}$ in the whole $H$ and $\theta$ range considered), the normalized $k_{p,11}/k_{p,11}^{max}$ is free from geometrical factors.

\begin{figure}[ht]
\centerline{\includegraphics[width=8.5cm]{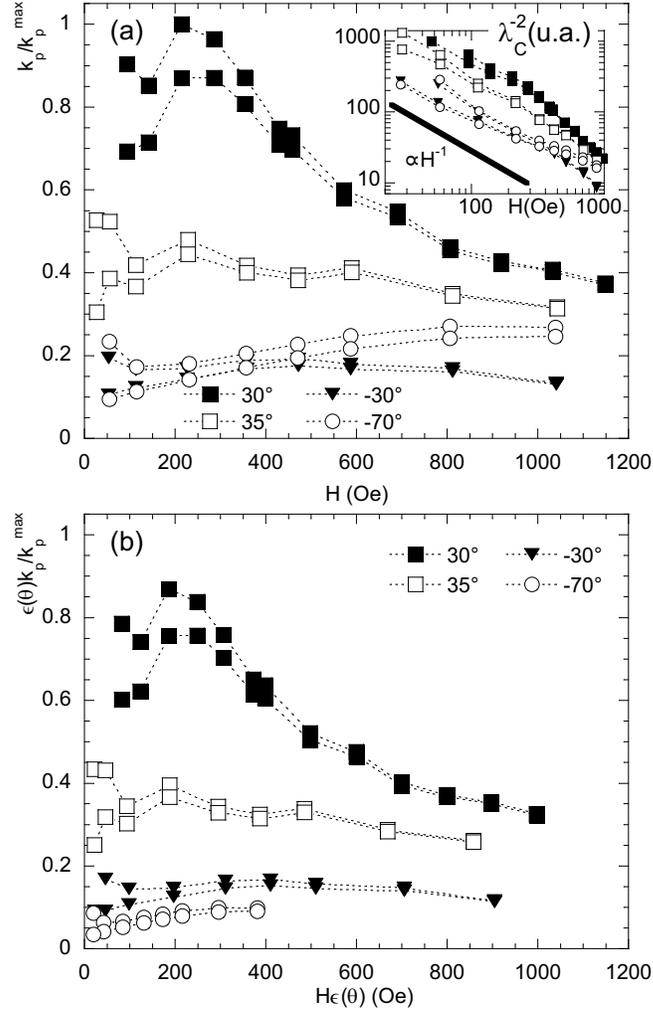}}
  \caption{Normalized pinning constant, as is (a) and scaled (b), as a function of d.c. applied field at selected angles. Inset of panel (a): $\lambda^{-2}_C$ vs $H$, with continuous line showing a $H^{-1}$ dependence for reference. Data digitized from reference \cite{Pasquini} as reported in the text.}
\label{fig:labusch}
\end{figure}

The result is reported in figure \ref{fig:labusch}a. The process of acquiring the $\lambda^2_R(B)$ data from the original plot, extrapolating $\lambda^2_R(B)$ to $B\rightarrow 0$ and subsequently inverting $\lambda_C^2(B)$, the latter being a quite small quantity for small fields especially for $\theta=30^\circ$ and $35^\circ$, is prone to a certain degree of uncertainty. Hence for each angle $\theta$ two curves are reported, delimiting the uncertainty area originated by the spread of the $\lambda^2_R(B\rightarrow 0)$ values. As a rough estimation of the spread, the range of values obtained by using a variable number ($2\div5$) of low field data points of $\lambda^2_R$ for the linear extrapolation to $B\rightarrow 0$ is used. 
For illustrative purposes, the so-obtained $\lambda^{-2}_C(B)$ is also shown (inset of figure \ref{fig:labusch}a), compared with the line representing a $H^{-1}$ field dependence which would yield a constant $k_p$. It should be noted that the larger uncertainty (confined to low fields) and the hump obtained at low fields for the $\theta=30^\circ$ data are of no consequence for the subsequent discussion, since the latter is focused on the data for $\theta=-30^\circ$ and $-70^\circ$, which are less uncertainty-prone due to their larger $\lambda^2_R$.

By inspecting figure \ref{fig:labusch}a, it can be seen that the pinning constant in the irradiated sample with $\vec{B}$ along the columnar defects ($\theta=30^\circ$) is larger than along the other directions. Moreover, by examining the curves with $\theta=-30^\circ$ and $-70^\circ$, it was noted \cite{Pasquini} that, despite being very similar at low fields, the two curves depart at near half the matching field $B_\Phi\approx350\;$Oe. Thus, it was inferred \cite{Pasquini} that the defects still determined an appreciable contribution to pinning even for the field tilted at $\theta=-30^\circ$, 60$^\circ$ far from the tracks directions. 
Actually, this observation can be further substantiated by resorting to the result of section \ref{sec:campbell}, namely equation \eref{eq:kptensor_Bdep}: were only point pinning present, the in-plane pinning constant at different fields and angles should scale so that by plotting $\epsilon(\theta) k_{p,11}$ vs $B\epsilon(\theta)$ all the curves should overlap. Any possible residual difference between the scaled curves should then be ascribed to the effects of correlated defects only, since the scaling removes the background effect of point pinning centers. The result of such scaling is reported in figure \ref{fig:labusch}b, where $\gamma=7$ has been used (taking \cite{poole} $\gamma=5\div8$ does not change the result). It can be seen that, even by having removed the effect of point pins, along the tracks directions pinning remains the largest. On the 
other hand, the absolute values for $\theta=-30^\circ$ are now larger than those for  $\theta=-70^\circ$. Since at these 
angles the contribution of twins and $a$-$b$ planes can be neglected (see also figure 2 in reference \cite{Pasquini}), one can infer that at $\theta=-30^\circ$ the artificial defects still reinforce pinning, yielding a higher pinning constant than the one measured at $\theta=-70^\circ$. This result further substantiates, on quantitative grounds, the observation done on qualitative grounds, i.e. based on the field dependence, reported in the original work \cite{Pasquini}.

\section{A.c. vortex motion resistivity}
\label{sec:highfrequency}

In this section I address the determination of the vortex motion resistivity tensor $\tens{\rho}_v$ in a general vortex motion regime in which both viscous drag and pinning effects are in action.
In order to introduce all the relevant quantities, I first recall the general models describing the high frequency vortex dynamics with reference to the commonly studied configuration of isotropic superconductors with $\vec{J}\perp\uvec{B}$ (section \ref{sec:scalarmodel}).
Second, I develop the full treatment for uniaxially anisotropic superconductors (section \ref{sec:acforceequation}). Then, some examples of data analysis based on the obtained results will be illustrated in section \ref{sec:rhovm_examples}.

\subsection{Short review of scalar models}
\label{sec:scalarmodel}

The scalar force equation describing the vortex motion in an isotropic superconductor with isotropic (point) pinning and $\vec{J}\perp\uvec{B}$ is, in the sinusoidal regime $e^{\rmi\omega t}$  \cite{GR,Golo}:
\begin{equation}
\label{eq:forceIso}
\eta v+\frac{k_p}{\rmi\omega}v=\Phi_0 J+F_{therm}
\end{equation}
where $F_{therm}$ is a stochastic thermal force causing thermal depinning (vortex creep). 
Different approaches  \cite{BrandtModel,CCiso}, with different ranges of applicability  \cite{PompeoPRB}, have been proposed to take into account creep effects. 
As an illustration, I follow here the description of thermal depinning in terms of the relaxation of the pinning constant $k_p(t)=k_p e^{-t/\tau_{th}}$ (reference \cite{BrandtModel}). The characteristic time for thermal activated depinning is:
\begin{equation}
\label{eq:tauth}
\tau_{th}=\tau_p e^{U/K_BT}
\end{equation}
where $\tau_p$ is the inverse of the (de)pinning angular frequency $\tau_p^{-1}=\omega_p=\eta/k_p$ (which will be commented on later) and $U$ is the activation energy. Equation \eref{eq:forceIso} can then be rewritten as  \cite{BrandtModel}:
\begin{equation}
\label{eq:forceIso1}
\eta v+\frac{k_p}{\rmi\omega}\frac{1}{1-\frac{\rmi}{\omega\tau_{th}}}v=\eta_\com v=\Phi_0 J
\end{equation}
where the complex viscosity $\eta_\com$ has been introduced:
\begin{equation}
\label{eq:etac0}
\eta_\com=\eta\left(1-\rmi\frac{\omega_p}{\omega}\frac{1}{1-\frac{\rmi}{\omega\tau_{th}}}\right)
\end{equation}
The corresponding scalar vortex motion resistivity is:
\begin{equation}
\label{eq:rhoBiso}
\rho_{v}=\frac{\Phi_0 B}{\eta_\com}=\frac{\Phi_0 B}{\eta}\frac{\varepsilon'+\rmi\frac{\omega}{\omega_0}}{1+\rmi\frac{\omega}{\omega_0}}
\end{equation}
where the characteristic angular frequency $\omega_0$ is:
\begin{equation}
\label{eq:omega0}
\omega_0=\tau_{th}^{-1}+\tau_p^{-1}
\end{equation}
and the creep factor $\varepsilon'$ is:
\begin{equation}
\label{eq:creep}
\varepsilon'=\frac{1}{1+e^\frac{U}{K_BT}}
\end{equation}
For $U\rightarrow\infty$ the creep is negligible, $\varepsilon'\rightarrow0$ and $\omega_0\rightarrow\omega_p$; consequently the vortex motion resistivity becomes: 
\begin{equation}
\label{eq:rhoGRiso}
\rho_{v}=\frac{\Phi_0 B}{\eta}\frac{1}{1-\rmi\frac{\omega_p}{\omega}}=\left(\rho_\ff^{-1}-\rmi\rho_C^{-1}\right)^{-1}
\end{equation}
This limit corresponds to the Gittleman\textendash{}Rosenblum (GR) model  \cite{GR}. From equation \eref{eq:rhoGRiso} it can be seen that the pinning angular frequency $\omega_p$ marks the transition between a ``low frequency'' and a ``high frequency'' regime: for $\omega\ll\omega_p$ the pinning force dominates over the viscous drag, yielding $\rho_{v}\rightarrow\rmi\rho_C$, while for $\omega\gg\omega_p$, a purely dissipative flux flow regime is recovered with $\rho_{v}\approx\rho_\ff$ yielding the same behaviour as in d.c. with no pinning.

Before concluding this short review, it is worth recalling the definition of the often used  dimensionless ratio $r$ \cite{JAP2009,rparameter1,tsuchiya,rparameter2}:
\begin{equation}
\label{eq:r}
r=\frac{\Im(\rho_{v})}{\Re(\rho_{v})}
\end{equation}
which, if creep is negligible (GR limit) yields:
\begin{equation}
\label{eq:rGR}
r=\frac{\omega_p}{\omega}=\frac{\rho_\ff}{\rho_C}
\end{equation}
The $r$ parameter can be directly computed from the complex resistivity $\rho_v$ and it is unaffected by any systematic error in the experiments. Physically, it allows to easily evaluate whether the vortex dynamics is in the pinning dominated ($r\gg1$) or flux flow dominated ($r\ll1$) regime.

\subsection{The anisotropic model}
\label{sec:acforceequation}

The force equation in anisotropic superconductors including both viscous drag and pinning, momentarily neglecting vortex creep (treated later), can be written as:
\begin{equation}
\label{eq:force3a}
\tens{\eta}\vec{v}+\frac{1}{\rmi\omega}\tens{k}_p\vec{v}=\Phi_0 \vec{J}\times\uvec{B}
\end{equation}
which, by defining the complex viscosity tensor as:
\begin{equation}
\label{eq:etac1}
\tens{\eta}_{\com}=\tens{\eta}-\rmi\frac{\tens{k}_p}{\omega}
\end{equation}
can be recast in:
\begin{equation}
\label{eq:force3b}
\tens{\eta}_\com\vec{v}=\Phi_0 \vec{J}\times\uvec{B}
\end{equation}
It is evident that the force equation \eref{eq:force3b} is formally equivalent to force equations written for the flux flow and Campbell regimes (equations \eref{eq:force1} and \eref{eq:force3}, respectively). Hence, it is straightforward to apply the previous results to the series of quantities $[\tens{\eta}_\com, \tens{\intr\eta}_\com, \tens{\intr\rho}_{v}, \tens{\intr\sigma}_{v}, \tens{\mu}_v, \tens{\rho}_{v}]$.
As a first result, within this framework one can write:
\begin{equation}
\label{eq:etac2}
\tens{\eta}_{\com}=-\xma{B}\tens{\intr\eta}_{\com}\xma{B}
\end{equation}
Equations \eref{eq:etac1} and \eref{eq:etac2} allow to write down:
\begin{equation}
\label{eq:etaci1}
\tens{\intr\eta}_{\com}=\tens{\intr\eta}-\rmi\frac{\tens{\intr k}_p}{\omega}\\
\end{equation}
Substituting equations \eref{eq:eta0i} and \eref{eq:kptensor_Bdep}  into \eref{eq:etaci1} yields:
\begin{align}
\label{eq:etaci2} 
\nonumber
\!\tens{\intr\eta}_{\com}\!(B, \theta)\!&=\!\left(\!\intr\eta_{11}(B,\theta)\!-\!\rmi \frac{\intr k_{p,11}(B,\theta))}{\omega}\!\right)\!\tens{M}^{-1}\!=\\ 
&=\left(1-\rmi\frac{\omega_p(B,\theta)}{\omega}\right)\tens{\intr\eta}(B,\theta)
\end{align}
It is evident that $\tens{\intr\eta}_{\com}$ inherits the angular dependencies and anisotropic properties from $\tens{\intr\eta}$ and $\tens{\intr k}_p$, represented by $\tens{M}$ and by the scaling law. 

I stress that, in obtaining equation \eref{eq:etaci2}, I have used a very important result for the analysis of the experiment, namely:
\begin{equation}
\label{eq:omegap}
\frac{{\intr k}_{p,ii}(B,\theta)}{\intr\eta_{ii}(B,\theta)}\!=\!\omega_p(B,\theta)=\omega_p(B\epsilon(\theta),0)
\end{equation}
which holds for $i=1..3$, i.e. for all the principal axes directions. The above equation shows that, contrary to the viscosity or the pinning constant, the pinning frequency is a scalar (using a GR model for the interpretation of the data, the same applies to the $r$ parameter). Together with the scaling property of $\omega_p$ highlighted in the last member, this fact implies that, whichever orientation is set for $\vec{B}$ and $\vec{J}$, the vortex system will always have the same pinning frequency at fixed $B/B_{c2}(\theta)$. This property suggest a straightforward method to check whether directional defects influence vortex motion. By plotting $\omega_p$ vs the scaled field $B/B_{c2}(\theta)$, any deviation from a scaling curve should indicate the influence of some directional effect other than the material anisotropy, since if extended pins are present equation \eref{eq:kptensor_Bdep} does not hold.

The material intrinsic conductivity and resistivity tensors are:
\begin{equation}
\label{eq:rho_vi1}
\tens{\intr\rho}_v=\left(\tens{\intr\sigma}_{v}\right)^{-1}=\Phi_0B \left(\tens{\intr\eta}_{\com}\right)^{-1}
\end{equation}
Using equation \eref{eq:etaci2}, the explicit expression for $\tens{\intr\rho}_v$ can be written down:
\begin{equation}
\label{eq:rho_vi2}
\tens{\intr\rho}_v(B,\theta)\!\!=\!\frac{\intr\rho_{\ff,11}\!(B,\theta)}{1\!-\rmi\frac{\omega_{p}(B,\theta)}{\omega} }\tens{M}\!=\!\intr\rho_{v,11}\!(B,\theta)\tens{M}\!=\!\intr\rho_{v,11}\!(B\epsilon(\theta),0)\tens{M}\!
\end{equation}
This is another important result of this work: similarly to $\tens{\intr\eta}_{\com}$,  $\tens{\intr\rho}_v$ retains the same anisotropy of the flux flow and pinning tensors, given by the mass anisotropy tensor $\tens{M}$ alone. Moreover, it satisfies the same angular scaling law, shown in the last equality, as $\tens{\intr\rho}_\ff$ and $\tens{\intr\rho}_C$.

It is important to remember that $\tens{\intr\rho}_v$ of equation \eref{eq:rho_vi2} is not directly measured, whereas the actually measured tensor is:
\begin{equation}
\label{eq:rho_v}
\tens{\rho}_{v}(B,\theta,\phi)=\intr\rho_{v,11}(B,\theta)\left[\frac{\tens{M}_B(\theta,\phi)}{\epsilon^2(\theta)}\right]
\end{equation}
From equation \eref{eq:rho_v} it can be noted that, contrary to $\tens{\intr\rho}_v$, the measured tensor $\tens{\rho}_{v}$ does not obey the angular scaling law, since it incorporates additional angular dependencies through $\tens{M}_B$.

Now I include the effects of flux creep. The pinning energy $U$ depends only on the magnetic field magnitude and direction and not on the direction of vortex motion. Moreover, it obeys the usual scaling law (with a constant scaling factor $\gamma^{-1}$ \cite{BGL}). Therefore the thermal depinning time $\tau_{th}$ (equation \eref{eq:tauth}) is a scalar $\omega_p=\tau_p^{-1}$. The pinning constant tensor of equation \eref{eq:kptensor_Bdep} can thus be modified to include creep as:
\begin{equation}
\label{eq:kpicreep}
\tens{\intr k}_p\!(B,\theta)\!=\!\intr k_{p,11}(B,\theta)\left(\!1\!-\frac{\rmi}{\omega\tau_{th}\!(B,\theta)}\right)\!\tens{M}^{-1}
\end{equation}
Consequently, the scalar vortex motion resistivity with flux creep, equation \eref{eq:rhoBiso}, can be generalized to the anisotropic case as follows, yielding an expression analogous to equation \eref{eq:rho_vi2}:
\begin{equation}
\label{eq:rho_vi3}
\tens{\intr\rho}_v(B,\theta)=\intr\rho_\ff\frac{\varepsilon'+\rmi\frac{\omega}{\omega_{0}}}{1+\rmi\frac{\omega}{\omega_{0}}}\tens{M}=\intr\rho_{v,11}(B,\theta)\tens{M}=\!\intr\rho_{v,11}\!(B\epsilon(\theta),0)\tens{M}\!
\end{equation}
where the characteristic frequency $\omega_0$ and the creep factor $\varepsilon'$ remain scalar values as in the isotropic case, reported in equations \eref{eq:omega0} and \eref{eq:creep}.
This property will prove important in the interpretation of the experiments.

It is worth stressing that the choice of the pinning constant relaxation as a model for flux creep \cite{BrandtModel} is not a limiting factor to the results obtained up to now: in fact, any pure thermal creep process (independent on the angle between the field and the current), possibly with a different definition of $\varepsilon'$ and $\omega_0$ (reference \cite{PompeoPRB}), would yield the same results.

\section{Application to experiments: the measured complex vortex resistivity in common setups}
\label{sec:rhovm_examples}

In this section I consider explicitly some typical experimental configurations and I derive specific expressions relating measured and material intrinsic quantities.
It should be noted that at microwaves the electromagnetic response of the superconductor is not only dictated by the motion of the vortices but arises from the coupling between the latter and the high frequency currents, which include both the normal and the superconducting components  \cite{CCiso,mgb2_omegap,fullTwoFluidMixedState}. Moreover, the actual physical quantity that can be directly measured is not the local complex resistivity, but the superconductor surface impedance \cite{rparameter1,tsuchiya,rparameter2,PRB1992,examplesZthin,examplesZ}. The incorporation of the present results in the calculation of the full electromagnetic response (in terms of the local complex resistivity and anisotropic surface impedance) is outside the scope of this paper and thus postponed to a future work.
In the following, the contribution of the coupled microwave currents will be neglected and the  superconducting material will be assumed to be in the form of thin film, a sample geometry widely used in microwave experiments \cite{examplesZthin}. In these conditions it can be shown that $\tens{\rho}_v$, being essentially proportional to the surface impedance tensor, can be taken as the actually measured quantity.

\subsection{Straight planar currents}
\label{sec:rhovm_examples1}
A straight a.c. current can be applied to flat thin films resorting to resonators having rectangular geometries \cite{PRB1992,rectangular}. Considering $\vec{J}\parallel\uvec{x}$ and using the resistivity tensor given by equation \eref{eq:rho_v}, the measured vortex resistivity is computed by applying equation \eref{eq:rhoJ}:
\begin{subequations}
\label{eq:rhoeff_exp4}
\begin{align}
\rho_{v}^{(x)}(B,\theta,\phi)&=\intr\rho_{v,11}(B,\theta)f_{L\phi}(\theta,\phi)\\
f_{L\phi}(\theta,\phi)&=\frac{\gamma^{-2}\sin^2\theta\sin^2\phi+\cos^2\theta}{\gamma^{-2}\sin^2\theta+\cos^2\theta} 
\end{align}
\end{subequations}
I recall that $\phi$ is the angle between the projection of the $\vec{B}$ field on the $x$-$y$ plane and the $x$ axis (figure \ref{fig:ref}). It can be seen that the effective, measured vortex motion resistivity consists in the product of two terms: the first is the resistivity $\intr\rho_{v,11}$ only, which in particular obeys the angular scaling law; the second one, denoted in the equation as $f_{L\phi}(\theta,\phi)$, is an additional angular dependence which arises from the Faraday and Lorentz actions. Consequently, as already anticipated commenting the whole tensor $\tens{\rho}_v$, the experimentally measured quantity does not obey the scaling law, and therefore care must be taken in isolating the intrinsic material property from the contribution given by the experimental setup before proceeding with the physical interpretation of the data.
Only in the case $\phi=\pi/2$ (i.e. the maximum Lorentz force configuration), $f_{L\phi}(\theta,\phi)=1$ and one has direct, experimental access to the intrinsic vortex resistivity $\intr\rho_{v,11}$.

\subsection{Rotational symmetric planar currents}
\label{sec:rhovm_examples2}
\begin{figure}[ht]
\centerline{\includegraphics[width=8.5cm]{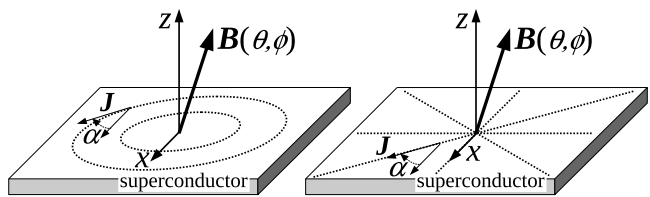}}
  \caption{ Circular (left panel) and radial (right panel) symmetric current patterns. $\alpha$ is the angle between the (local) current density $\vec{J}$ and the $x$ axis.}
\label{fig:rotcurrents}
\end{figure}

A rotational symmetric planar geometry is often used for measurements of the vortex-state microwave response. Examples are cylindrical resonators \cite{tsuchiya,cylres}, in which (using cavity perturbation techniques \cite{perturbation}) the superconducting sample is located on the circular bases, and the so-called Corbino disk setup \cite{creepCorbino,JSNM2006,Corbino}, in which the superconducting sample short-circuits an open-ended coaxial cable. In both cases, rotational symmetric currents (circular and radial for the resonator and the Corbino disk, respectively) are induced along the superconductor $a$-$b$ plane (see figure \ref{fig:rotcurrents}), in the frequent case in which the superconductor $c$-axis is perpendicular to its surface.
The actually measured \cite{APL2013p,circurrent} resistivity comes out as an angular average \cite{Collin} over the current pattern:
\begin{align}
\label{eq:rhoeff_exp51}
\nonumber
\rho_{v}^{(\circ)}&(B,\theta)=\frac{1}{2\pi}\int_0^{2\pi}\left(\tens{\rho}_v(B,\theta,\phi)\uvec{J}(\alpha)\right)\cdot\uvec{J}(\alpha) d\alpha=\\
=&\frac{1}{2\pi}\frac{\rho_{v,11}(B,\theta)}{\epsilon^2(\theta)}\int_0^{2\pi}\left(\tens{M}_B(\theta,\phi)\uvec{J}(\alpha)\right)\cdot\uvec{J}(\alpha) d\alpha
\end{align}
where $\alpha$ is the angle between the (local) current density $\vec{J}$ and the $x$ axis (see figure \ref{fig:rotcurrents}).
Exploiting the uniaxial anisotropy together with the circular symmetry of the current pattern, the computation can be equivalently and more simply done by averaging equation \eref{eq:rhoeff_exp4} over all possible values of $\phi$:
\begin{align}
\nonumber
\rho_{v}^{(\circ)}(B,\theta)&=\frac{1}{2\pi}\int_0^{2\pi}\rho_{v}^{(x)}(\theta,\phi) d\phi
\end{align}
The result is:
\begin{subequations}
\label{eq:rhoeff_exp53}
\begin{align}
\label{eq:rhoeff_exp53a}
\rho_{v}^{(\circ)}(B,\theta)&=\intr\rho_{v,11}(B,\theta)f_{L}(\theta)\\
f_{L}(\theta)&=\frac{\frac{1}{2}\gamma^{-2}\sin^2\theta+\cos^2\theta}{\gamma^{-2}\sin^2\theta+\cos^2\theta}
\end{align}
\end{subequations}
where $f_{L}(\theta)$ is the average of $f_{L\phi}$ over all the $\phi$ values.
The same comments proposed for equation \eref{eq:rhoeff_exp4} hold also here. Moreover one can note that, in the present case, if $\intr\rho_{v,11}\propto(B/B_{c2}(\theta))^\beta$ equation \eref{eq:rhoeff_exp53} still yields a scaling law \cite{APL2013p}. However, in the interpretation of the experiments one has to be careful and not confuse this artificial scaling function with the theoretical scaling expression.

\subsection{Angle-dependent effective quantities}
It is interesting to note that a typical GR model analysis in a isotropic superconductor extracts the vortex parameters $\rho_\ff$, $r$ and $k_p$ from the complex measured $\rho_{v}$ of equation \eref{eq:rhoGRiso} as follows:
\begin{align*}
\label{eq:rhoGR2}
  \rho_{\ff} & =\Re(\rho_{v}) \left[1+\left(\frac{\Im(\rho_{v})}{\Re(\rho_{v})}\right)^2\right] \\
  r &=\frac{\Im(\rho_{v})}{\Re(\rho_{v})} \\
  k_p &=\frac{r}{\rho_\ff}\omega B \Phi_0 = \omega B \Phi_0 \frac{\Im(\rho_{v})}{\Re^2(\rho_{v})+\Im^2(\rho_{v})}
\end{align*}
On the other hand, when performing measurements on an anisotropic superconductor probed with rotational symmetric current patterns leading to equation \eref{eq:rhoeff_exp53}, this computation would yield the following {\em effective} quantities (apart from the additional $\phi$-dependence, the same holds for the straight current setup of equation \eref{eq:rhoeff_exp4}):
\begin{subequations}
\label{eq:rhoGR3}
\begin{align}
  \rho_{\ff,\eff}(B,\theta) & =\intr\rho_{\ff,11}(B,\theta) f_{L}(\theta) \\
  r_{\eff}(B,\theta) &=r(B,\theta) \\
  k_{p,\eff}(B,\theta) &=\frac{\intr k_{p,11}(B,\theta)}{f_{L}(\theta)}
\end{align}
\end{subequations}
It can be seen that the parameter $r=\omega_p/\omega$ is directly obtained from the measured quantities: this is an interesting result, which allows a direct evaluation of the material anisotropy of the system without the need to deal with Lorentz-dependent contribution $f_{L}(\theta)$. On the other hand, both $\rho_{\ff,\eff}$ and $k_{p,\eff}$ show an additional angular dependence through $f_{L}(\theta)$. Therefore, in the analysis of angular data care must be devoted in correctly extracting the intrinsic quantities instead of the effective ones: this requires to evaluate in some way the $f_{L}(\theta)$ function, which in turn requires the knowledge of the anisotropy factor $\gamma$.

Further comments can be done considering a scaling analysis performed starting from the effective quantities of equation \eref{eq:rhoGR3}. 
Once the intrinsic quantities $\intr\rho_{\ff,11}$ and $r$ are extracted, they can be checked against the scaling prescription. If $\intr\rho_{\ff,11}$ is found to satisfy the scaling, and in the same time $r$ or, equivalently $\intr\rho_{C,11}$, is not, this result would unambiguously indicate the presence of directional pinning contributions such as extended defects. 
This type of analysis is performed in references \cite{APL2013p}, where it enabled the accurate extraction of the intrinsic anisotropy of BaZrO$_3$ added YBa$_2$Cu$_3$O$_{7-\delta}$ thin films, the unambiguous identification in the angular dependent pinning constant of extended pinning acting effectively at microwave regimes, and the identification of a Mott-insulator effect through a comparative study with d.c. $J_c$ measurements.

\section{Summary}
% \label{sec:summary}
The electrodynamics model for transport measurements in the mixed state in uniaxial anisotropic superconductors has been discussed in various regimes, namely the dissipative free flux flow regime, the dissipationless low frequency pinned Campbell regime and the high frequency regime, where dissipation and pinning effects are comparable and thermal depinning/creep appears.
Arbitrary orientations between the applied field, the applied current and the anisotropy axis have been considered. 
Vortex parameters, like the viscous drag, the vortex mobility and pinning constant, have been derived in tensor form for arbitrary field orientations. It has been shown that the tensors describing point pinning share the same structure of the flux flow tensors, and that the measured quantities, differently from the corresponding intrinsic quantities, in general do not satisfy the angular scaling laws.
A full tensor model for the a.c. vortex motion resistivity, including creep, pinning and flux flow, has been presented.
Relations between experimentally measured quantities and intrinsic material properties have been given, showing that care must be put in separating the material intrinsic angular dependence from the one arising from the geometry of the setup. 
Examples of data analysis based on the results obtained have been provided and discussed, showing how to identify the various contributions to the measured angular dependencies, consisting in intrinsic (charge carrier mass) anisotropy, extrinsic  preferential orientations (such as those due to extended defects) and experimental geometry contributions arising from the Faraday and Lorentz actions.

Further possible developments include the use of the present full tensor model of vortex motion in the high frequency/microwave regime where a coupling between microwave currents and vortex motion occurs and the experimentally measured quantity is the (anisotropic) surface impedance. Another interesting extension is the determination of the pinning constant tensor when both point pins and extended defects are present.

\ack
This work has been supported by Regione Lazio and partially supported by the Italian FIRB project ``SURE:ARTYST'' and by EURATOM.

The author warmly acknowledges the helpful discussion with prof. E. Silva.
\\

\end{document}